\documentclass[12pt,twocolumn]{article}
\usepackage[a4paper,left=20mm,right=20mm,top=25mm,bottom=25mm,includeheadfoot]{geometry}

\usepackage{mathpazo} 
\usepackage{graphicx}
\usepackage{amsmath,textcomp}
\usepackage{makeidx}
\makeindex

\usepackage{sectsty}
	\allsectionsfont{\sffamily\raggedright}
	\sectionfont{\sffamily\large\raggedright}
	\subsectionfont{\sffamily\normalsize\raggedright}
\usepackage{cite} %%%%%%%%%%%%%%%%%%%%%%%%%%%%

\usepackage[caption=false]{subfig}% Support for small, `sub' figures and tables %%%%%%%

\usepackage{widetext}
\usepackage{flushend}
\usepackage{cuted}
\usepackage{color}
\usepackage{graphicx}
\usepackage{hyperref}
\usepackage{pstricks,amssymb}

\usepackage{fancyhdr}
\begin{document}
%Do not change the \vspace command, which is needed to suppress extra space above the title.
\title{\vspace{-2em}\bfseries\sffamily Marbles and Bottles-or Boxes illustrate Irreversibility and Recurrence}
\author{\normalsize Anna Maria Aloisi${^1}$ and Pier Franco Nali${^2}$\\[2ex]
$^{1}$IPSIA A. Meucci, Cagliari, Italy (Ret.).\\
{\tt annamaria.aloisi@istruzione.it}\\[2ex]
$^{2}$Via Tempio 29, 09127 Cagliari, Italy.\\
{\tt ampfn@tiscali.it}
}

\date{\itshape Submitted on dd-MMM-yyyy}

\maketitle

\thispagestyle{fancy}

%Do not change the \sffamily command, it is needed to ensure that the abstract appears in a different font.
\begin{abstract}
{\sffamily
Imaginary assemblies of intercommunicating bottles-or boxes, in which marbles circulate performing random walks, can help modeling (and hence understanding) slow and long-lasting diffusion processes and enable easy evaluations of typical times involved in their dynamics. Compared to the more traditional approaches to explaining diffusion inspired by the Ehrenfest\textquotesingle s urn problem, these simple random walk models offer a more straightforward way to illustrate irreversibility\slash recurrence issues. 
}\\
\hrule
%Do not change the \hrule command, it is needed to separate the abstract form the main text.
\end{abstract}

\section{Introduction} \label{intro}
Ideal models based on ``Marbles and Boxes'' are well suited to acquaint students with the basics of the diffusion and transport dynamics in gases and other media. Mostly implemented by computer algorithms, these abstract artifacts can sometimes have a material counterpart suitable for developing hands-on classroom activities. An example is the ``Marble Game'', \footnote{Not to be confused with the popular LEGO Marble Maze Game (Labyrinth).} proposed some years ago in the U.S. in connection with the reforms of the STEM curriculum \cite{Nelson}. In this game \textit{N} marbles are distributed between two boxes. Marbles jump between boxes (in both directions) at a constant rate, with a random extraction rule (e.g., rolling a multi-sided die) that decides which marble will next jump to the other box: if the rolled number is less than or equal to the number of marbles in \textit{box1}, then a marble is moved from \textit{box1} to \textit{box2}, otherwise a marble is moved from \textit{box2} to \textit{box1}.

The modeling with the ``Marble Game'' is a kinetic Monte Carlo (kMC) simulation of the classic Ehrenfest\textquotesingle s two-urn model system (also known as dog-fleas model) and can be physically realized if the number \textit{N} of marbles is not too large (say \(N=10\)). This allows the possibility for a classroom activity with a physical representation of the game, on which the students can play moves by hand according to set rules, and gives support in a sensory form to the idea that diffusion and molecular transport phenomena are marked by the evolution of a dynamical system. Once the process has started, its evolution is examined through the concepts of probability, statistical equilibrium and randomness of the states reached by the system. By playing the game, and at a later time running computer simulations, students discover that equilibrium is the result of a dynamical process yielding to an irreversible tendency, and recognize the key role that randomness plays in the diffusion mechanism (ruled by Fick\textquotesingle s first law) modeled by the game.  

In this paper, we consider a ``Marbles and Bottles-or Boxes'' model explaining diffusion, reported in the 90's by Clifford Pickover, a researcher at IBM Watson Research Center, who first introduced it in form of a math puzzle connecting time to probability \cite{Pick01,Pick02}. This model - that is as a matter of fact the well-known random walk in one dimension - leads to a startling representation of time intervals in terms of solutions to simple random walk and diffusion problems, and offer an alternate view to the more classical approaches, based on the ``Marble Game'' or similar models, to explain macroscopic irreversibility. 

The structure of this paper is organized as follows. In section \ref{sec2}, we outline how the ``Marbles and Bottles-or Boxes'' model operates and what kind of physical systems it could represent. In section \ref{sec3}, we introduce some very elementary concepts about random walks and stochastic processes, required in order to appreciate the statistical arguments developed in subsequent sections. In sections \ref{sec4}, \ref{sec5}, and \ref{sec6}, we present three sample applications of the above concepts and offer a proof, only relying on elementary probability theory and finite difference equations, of a formula to calculate the first-passage time through a generic node of a linear chain assembly. In sections \ref{sec7} and \ref{sec8}, in order to illustrate irreversibility\slash recurrence issues, we introduce a variant of the basic random walk model applicable to various physical situations and briefly recapitulate the Ehrenfest\textquotesingle s urn model along with generalizations thereof. Some remarks on the discrepancy between recurrence and irreversibility at the macroscopic level are developed in sections \ref {sec9} and \ref{sec10}, where simple simulation results are also summarized. In section \ref{sec11} a gas effusion process is presented as a microscopic analogue of the model outlined in section \ref{sec2}. Conclusions are drawn in section \ref {sec12}, along with a brief discussion of the pedagogical value of the above-mentioned models and their limitations.

\section{Time in a bottle-or a box} \label{sec2}
What happens if we place a marble into a large glass bottle that had a small opening at its top and we began to shake the bottle randomly? How long would it take for the marble to leave the bottle through the tiny aperture if we were to continue shaking? The average time for the marble to get out of the bottle will obviously depend on the size of the hole (and the size of the marble). Let us say the marble popped out after 1 hour of constant bouncing around in the bottle.

What would happen now if we were to place a series of bottles together so that only a small opening connected the bottles: Bottle 1 connects to Bottle 2, Bottle 2 connects to both Bottle 1 and Bottle 3 (see figure \ref{fig:image01}), and so on. Nothing prevents us from imagining that more and more bottles could be connected using their small openings, all in the same way to make up a linear assembled chain. The last bottle of the sequence (say Bottle \textit{n}) opens to the outside world. (Assume this is an ideal system: it has no friction, gravity, etc.). How long would it take for the marble to exit Bottle \textit{n}? We must bear in mind that in each of the intermediate bottles the marble, in its random motion, has just as likely a chance of moving into a previous bottle as it does moving forward. Let us also assume that it takes one hour for the marble to find an opening as it did in the single-bottle experiment.\footnote{All models involve some simplification. Models as those considered in this paper, in which imaginary marbles of a proper size are placed in imaginary boxes, are well suited for high-speed computer simulations and were used in Physics since the pioneering work by Berni Adler \cite{Adler} on hard sphere systems.} 

It can be easily shown, virtually starting from scratch, that the average bottle number reached in a given time approaches a constant. This suggests a method for drawing diagrams, of immediate visual effectiveness for a student, connecting bottles-or boxes (or chambers or other kind of container) in long chains representing large expanses of time, in fact so large that the flow of marbles through them appears as essentially irreversible.

In addition, the imaginary assembly of bottles just described could represent a macroscopic model demonstrating the diffusion of an extremely rarefied gas in a network of high vacuum flasks connected by highly selective porous seals.

Finally, it could be used as a simulation tool suitable to give just an idea of the difference between a macroscopic and a microscopic system: how fast (and how small) should the marble be, in order to reduce the ``diffusion'' time to values that can actually be experienced even for this kind of relatively complex system?

\section{Random walks and Poisson processes} \label{sec3}

The concept of random walk (or drunk man walking) was introduced by the English philosopher and statistician Karl Pearson (1857-1936) in a letter to Nature, dated July 27, 1905 \cite{Pearson}, where the following problem was presented for the first time: \textit{``A man starts from a point O and walks l yards in a straight line; he then turns through any angle whatever and walks another l yards in a second straight line. He repeats this process n times. I require the probability that after these n stretches he is at a distance between r and r + dr from his starting point, O.''}.

Intensively studied in the 20\textsuperscript{th} century, examples of random walk are ubiquitous, from Brownian motion to diffusion processes in chemical-physics, biology and sociology \cite{Daboni}; other examples, frequently quoted, concern the motion of a body through a series of adjacent passages or, occasionally, in confined geometries, with various applications to transport and separation processes \cite{Weiss01,Weiss02,Saltzman,Burada,Xu,DasGupta}.  As a special kind of stochastic process, the random walk is a mathematical model that schematizes, using probabilistic-statistical methods, the time course of a random phenomenon. 

In the following, it is assumed that the search for the exit in the random motion of the marble inside a bottle is a stochastic process, more specifically a Poisson process, in which events occur in time completely at random at intermittent times, like incoming calls to a telephone. The mathematical description of natural phenomena such as radioactive disintegration, and of lots of demographic, economic and industrial production processes are based on the Poisson model\cite{Daboni,Weiss01,Weiss02,DasGupta}.

In the Poisson model the probability $\Delta{\textit{p}}$ for the marble to get to the opening in a time interval $\Delta{\textit{t}}$ is proportional to same $\Delta{\textit{t}}$, namely  $\Delta{\textit{p}}=\lambda\Delta{\textit{t}}$, where $\lambda$ is a constant whose dimensions are the inverse of time and represents the probability per unit time for the marble to leave Bottle 1. It can be shown that the average time the marble would take to find the opening is $\lambda^{-1}$.\footnote{More exactly, it can be shown that the number of openings found in time \textit{t} is, on average, $\mu =\lambda t$ (See, e.g., \cite{Daboni,DasGupta}). For  $\mu =1$ it follows the assertion.} From now on, we will denote the above quantity by $m_1$ (1 hour in our example), which corresponds to the time step in the random walk. If the bottle includes more openings, the exit probability increases according to their number $M$ (i.e., $\Delta{\textit{p}}=M\lambda\Delta{\textit{t}}$), while the average time for the marble to get out follows the inverse proportion, namely $(M\lambda)^{-1}=m_1\slash M$.

Very simple statistical reasonings allow us to analyze the motion of the marble within the bottle chain and derive an equation for the average time Bottle \textit{n} is exited for the first time.\footnote{In the stochastic jargon this is defined as first-passage time or hitting time.} The above problem is widely and well-known in Statistics, and reduces to the simple random walk on $\mathbb{N}$ (the set of natural numbers) of length \textit{n}. As we do not require the reader to have any prior knowledge on random walk processes, our analysis was conducted with the intent of achieving the goal by steps, and to this end it took into consideration three cases:
\begin{enumerate}
\item Only one free-flowing connection between the \textit{n} bottles (this reduces to the trivial case of unidimensional diffusion of a free walker).
\item 	The addition of backward openings (case of a biased walker with equal step size in both directions but unequal probability of taking forward and backward paths).
\item 	Addition of backward openings, after considering that the first bottle is different from the rest (due to the reflecting barrier at the starting node).
\end{enumerate} Further generalizations could be imagined considering more complex network topologies constituted by nodes having connections through both forward and backward paths in various combinations of step sizes and direction probabilities. 

\section{Random walk and diffusion in a linear chain assembly} \label{sec4}

In the simplest case (1.) remember from section \ref{sec2} that in each of the intermediate bottles (none at the end of the chain assembly) the marble has just as likely a chance of passing to the next bottle as it does to regressing into a previous one. 

Let $m_k$ represent the expected amount of time the marble would take to pass from Bottle \textit{k} \,to Bottle $k+1$ (including any regressions to previous bottles). Then, for $k>1$, there is a 50 percent chance that the marble will go directly from Bottle \textit{k} \,to Bottle $k+1$, incurring an average time $m_1$, and a 50 percent chance that it would regress to Bottle $k-1$, in which case the average time to return to Bottle \textit{k} \,and then to move to Bottle $k+1$ would be $m_{k-1}+m_k$. This leads to the following difference\footnote{Following a widely used definition, the term difference equation is treated here as synonymous with recurrence relation.} equation involving the averages of stochastic variables:
\begin{equation} \label{eq:eps1}
m_k=\frac{1}{2}m_1+\frac{1}{2} (m_{k-1}+m_k ),
\end{equation}
which simplifies to $m_k=m_{k-1}+m_1$. 
By induction, its solution is: $m_k=km_1$. 

So, we get the simple but interesting result that the average time to pass from Bottle \textit{k} \,to Bottle $k+1$ is \textit{k} times the expected time to exit Bottle 1. The average time \textit{m} \,to move from Bottle 1 to Bottle \textit{n} \,would be $1+2+\ldots+(n-1)=n(n-1)\slash 2$ times the average time $m_1$  to find an opening (For a large number of bottles the latter equation might be approximated by $n^2\slash 2$). We ignore here the fact that the first bottle is different from the rest of the chain as well as any complications depending upon whether or not continuous space or discrete space settings were assumed. Depending on the case, the results are slightly different when the assemblage contains just a few chambers. However, as stated, if a large number of chambers are considered, \textit{m} increases according to $n^2$. It can be shown, in a strict form, that the first-passage probability for the discrete random walk and the continuum diffusion in transmission mode are asymptotically identical \cite{Redner}.

According to the stochastic jargon, the free state of the marble is an absorbing state (the state in which the marble is removed from the system) and the \textit{m}  quantity is the average absorption (or escape) time. The chain has a reflecting node (supposed at the origin) and ends on an absorbing barrier. At the start of the experiment, the marble is placed in the reflecting node. It should be noted that if the connections between the bottles were equipped with valves preventing the marble from making regression, the average time to traverse the chain and reach the absorbing (free) state would be the viable minimum, equal to \textit{n} \,times $m_1$.

\section{Random walk with addition of backward connectors} \label{sec5}

Let us now examine the case (2.), for $k > 1$, with $M+1$ possible exits from a bottle (except Bottle 1), one forward and \textit{M} backward (see figure \ref{fig:image02}). In other words, the system is ruled so that one opening is free flowing, while the remaining have a one-way valve allowing the marble to only travel in a backward direction (a direction away from the opening of final egress).

Assuming as before that finding an exit is a Poisson process, the average time to find the first of $M+1$ exits is  $m_1\slash (M+1)$. There is a probability $1\slash (M+1)$ that this exit will be forward, in which case no additional transition time is required. Also, there is a probability $M\slash (M+1)$ that the first exit will regress to Bottle $k - 1$, in which case the additional average time to return to Bottle \textit{k} \,and then to progress to Bottle $k + 1$ would be $m_{k-1}+m_k$. This leads to the modified difference equation:
\begin{equation} \label{eq:eps2}
m_k=\frac{1}{M+1} m_1+\frac{M}{M+1}(m_{k-1}+m_k ),				 
\end{equation} (remember that eq. \eqref{eq:eps2} is valid for $k>1$), which simplifies to $m_k=Mm_{k-1}+m_1$. Its solution (verifiable by induction) is:

%\begin{widetext}
\begin{equation} \label{al3}
\begin{split}
m_k & =(\textstyle\sum_{j=0}^{k-1}M^j ) m_1{} \\
{} & =\left\{ \begin{array}{ll}k m_1, &  \textrm{if $M=1$,}\\\displaystyle\frac{M^k-1}{M-1} m_1, & \textrm{if $M>1$.}\end{array} \right.
\end{split}
\end{equation}
%\end{widetext}
(Note that eq. \eqref{al3} is also valid for $k=1$).
The average time to exit the \textit{n}\textsuperscript{th} bottle (i.e. the average absorption time of the chain) is thus:

\begin{widetext}
\begin{align} \label{al4}
m=\textstyle\sum_{k=1}^n m_k=
\left\{ \begin{array}{ll}[n(n+1)\slash 2] m_1, & \textrm{if $M=1$,}\\\displaystyle\frac{M^{n+1}-M-n(M-1)}{(M-1)^2} m_1, & \textrm{if $M>1$.}\end{array} \right.
\end{align}
\end{widetext}
The second line in the equation array \eqref{al4} might be approximated by 
\begin{align} \label{al5}
m \approx \frac{M^{n+1}}{(M-1)^2} m_1 \end{align} for large \textit{n} (say $n \geq 10$).

\section{Simple random walk model} \label{sec6}

We have assumed that the rate of exit from each hole is the same and hence the rate of exit from a bottle increases according to the number of holes, namely it is $M+1$ times the rate of exit from Bottle 1, which has only one hole (this makes the first bottle different from the rest). Then the marble will find an exit with a probability per unit time ${\Delta p\slash\Delta t=(M+1) \lambda=(M+1) \slash m_1}$, incurring an average time  $m_1\slash (M+1)$.

If it is assumed (case 3.) that the average time to leave a bottle is the same for all bottles including the first (the simple random walk model), then the constant in the difference equation (probability of leaving Bottle 1) will increase by a factor of $M+1$. This can be obtained, for example, by enlarging the hole of Bottle 1 by the same factor. Then, relative to the first bottle, the remaining will have an exit probability $1\slash(M+1)$ for each opening. It follows that the marble will take directly the forward exit with a probability per unit time ${\Delta p\slash\Delta t=\lambda\slash(M+1)=1\slash[m_1 (M+1)]}$ incurring an average time $(M+1) m_1$. Substituting in equation \eqref{eq:eps2} the constant $m_1$ with $(M+1)   m_1$ the solution then changes to: 

\begin{widetext}
\begin{align} \label{al6}
m_k=(2\textstyle\sum_{j=0}^{k-1}M^j -1) m_1=
\left\{ \begin{array}{ll}(2k-1) m_1, & \textrm{if $M=1$,}\\\displaystyle\frac{2M^k-M-1}{M-1} m_1, & \textrm{if $M>1$.}\end{array} \right.
\end{align}
%\end{widetext}
The average time required to get past the \textit{n}\textsuperscript{th} bottle becomes:

%\begin{widetext}
\begin{align} \label{al7}
m=\textstyle\sum_{k=1}^n m_k=\left\{ \begin{array}{ll}n^2 m_1, & \textrm{if $M=1$,}\\\displaystyle\frac{2(M^{n+1}-M)-n(M^2-1)}{(M-1)^2} m_1, & \textrm{if $M>1$.}\end{array} \right.
\end{align} 
%\end{widetext}
The second line in the equation array \eqref{al7} might be approximated by
\begin{equation}\label{eq:eps2.1}
m \approx \frac{2 M^{n+1}}{(M-1)^2} m_1 \end{equation} (for large $n$).
\end{widetext}

The latter is the approximate equation reported by Pickover in \cite{Pick01,Pick02}. It was derived in 1991 by Shriram Biyani, Pickover\textquotesingle s colleague at IBM, using the statistical arguments exposed above \cite{Pick03}. Various other approaches to first-passage problems on $\mathbb{N}$ have been devised, and some of them can be found on the Internet (see, among others, \cite{site1}). The number of steps to go from $0$ to \textit{n} in a simple random walk on the line of natural numbers, with probability $p$ in the forward direction (except at the starting position, where the probability is $1$) and with probability $1-p$ in the opposite direction, is  expressed as 
\begin{align} \nonumber
\frac{n}{2p-1}+\frac{2p(1-p)}{(1-2p)^2}\left(\left(\frac{1-p}{p}\right)^n-1\right)\textrm{,}
\end{align} which reduces to the second line in the equation array \eqref{al7} for $p=1\slash (M+1)$ and $M>1$. 

We preferred proposing here the Biyani\textquotesingle s derivation  because of its greater simplicity compared to other approaches. So, equation \eqref{eq:eps2.1} (as its avatar \eqref{al5} reported in the previous section) may be used for $n \gg 1$ and $M>1$ - that is for large chains in the presence of one-way backward connectors in addition to the free flowing connection - and gives the average time until the \textit{n}\textsuperscript{th} bottle is exited for the first time (first-passage or hitting time). As stated in section \ref{sec2}, it turns out that the average bottle number reached in a given elapsed time approaches a constant.

\section{Random walk with restarts: The Sisyphus force} \label{sec7}
Many variants of the simple random walk model could be developed as the bottles can connect to each other through thin tubes in lots of combinations, scaling up the connections from simple linear chains to complex networks. Among the most interesting, the introduction of one-way backward connectors directly to Bottle 1 (figure \ref{fig:image03}), bypassing the rest of the chain. This is a special case, susceptible of a simplified mathematical treatment, of the random walk with random restarts, in which, starting at the origin, the walker faces two choices: either moving forward to the next node, or jumping back to the starting node with a ``restarting probability'' \textit{r},\footnote{In the random walk with restarts (RWR) standard algorithm, the walker is allowed to move to a randomly chosen neighbor (with a certain probability $p=1-r$), or to jump back to the origin with probability \textit{r}.} acting like a restoring force that will tend to bring the walker back to origin. This force is known as ``Sisyphus force'', and the walking process as ``Sisyphus random walk'' \cite{Montero}, by analogy with the Greek myth of Sisyphus. The effect considered underlies, in the context of Doppler laser cooling, the behavior of some physical mechanisms by which alkali atoms climbing from the ground level to higher excited states experience an increasing probability of being optically pumped into a minimum potential energy
state from where the process restarts. 

In this situation, if the marble experiences a regression, suddenly going back to Bottle 1 through any of the additional tubes, then the process restarts afresh from the source node, and the marble must re-cross all intermediate nodes in order to return eventually to Bottle \textit{k}. This is expressed by the equation: 

\begin{widetext}
\begin{equation} \label{eq:eps3}
m_k=\frac{1}{M+1}m_1+\frac{M}{M+1}(m_1+\ldots+m_{k-1}+m_k ),
\end{equation}
\end{widetext}
(valid for $k>1$), which simplifies to \begin{displaymath} m_k=m_1+M(m_1+m_2+\ldots+m_{k-1}).\end{displaymath} 
Hence, it is obtained, by induction,
%\begin{widetext}
%\begin{eqnarray*}
\begin{equation} \nonumber
\begin{split}
m_{k+1} & =m_1+M(m_1+\ldots+m_{k-1}+m_k) \\ 
& =m_1+M(m_1+\ldots+m_{k-1})+Mm_k \\
& =(M+1)m_k\textrm{,}
\end{split}
%\end{eqnarray*}
\end{equation}
%\end{widetext}
and at last $m_k=(M+1)^{k-1} m_1$, valid for $M\geq 1$. ($M=1$ this time is not a special case). 
The average absorption time is: 

%\begin{widetext}
\begin{equation} \label{al10}
\begin{split}
m=\sum_{k=1}^n m_k & =\frac{(M+1)^n-1}{M} m_1 \\
& \approx \frac{(M+1)^n}{M} m_1
\end{split}
\end{equation}
%\begin{align} \label{al10}
%m=\sum_{k=1}^n m_k=\frac{(M+1)^n-1}{M} m_1 \approx \frac{(M+1)^n}{M} m_1
%\end{align}
(for large $n$). 
%\end{widetext}
Assuming the same average exit time for all bottles, including the first one, we have to replace in the equation \eqref{eq:eps3} the constant $m_1$ by ${(M+1) m_1}$ and make the necessary simplifications\footnote{The simplifications are straightforward and are left as an exercise for the more energetic appenders.}, thus getting 

\begin{equation} \label{al11}
\begin{split}
m & =\frac{(M+1)^n-1}{M}(M+1)m_1 \\
& \approx \frac{(M+1)^n}{M}(M+1)m_1
\end{split}
\end{equation}
(for large $n$).

All the former results might be further generalized for a generic \textit{p} probability for the marble to directly take the forward exit (and a generic restarting probability $r$) by replacing $M=1\slash p-1=r\slash p$ in the expressions given above.\footnote{$m=\frac{p^{-n}-1}{1-p}m_1\approx \frac{p^{-n}}{1-p}m_1$. (Cfr. eq. (18) in ref. \cite{Montero}).}

\section{Ehrenfest\textquotesingle s urn experiment and recurrence time} \label{sec8}

The Marble Game we encountered in section \ref{intro} is a kMC simulation of the Ehrenfest\textquotesingle s urn experiment. This thought experiment, originally conceived in the wake of the grand debate about the apparent contradiction between second law of thermodynamics and Boltzmann's kinetic theory of gases, was authoritatively defined by Kac ``one of the most instructive models in the whole of physics'' \cite{Kac59}. In the simplest version of the experiment \cite{Ehrenfest}, one starts with \textit{N} marbles in the left urn (or \textit{urn0}). The marbles are numbered from one to \textit{N} and a third urn exists, containing \textit{N} cards with a natural number from one to \textit{N} printed on each. The procedure considers drawing a card at random from the cards-urn, looking at the number printed on it, drawing the marble with the corresponding number from its urn, and putting it in the other urn. (The cards are returned to the card-urn after each observation). The procedure is carried out repeatedly, virtually endlessly. Then, simply relying on the counting of the possible configurations and forcing the process to be repeated on a fairly large number of moves, we expect the initial state with the whole of the \textit{N} marbles in the \textit{urn0} to be reproduced with a probability of $2^{-N}$. 

The literature also reports a number of interesting multi-urn extensions\slash modifications of the original Ehrenfest\textquotesingle s dog fleas model, which would be worth exploring \cite{Kao,Luan,Clark,Song,Guemez,Banta,Klein}. In latter models, \textit{N} marbles circulate in the network relying on either directed or random mechanisms that distribute the marbles until some specific condition is reached, or the first-passage to some special state is detected.

In the case of a multi-urn experiment (\textit{M+1} urns from \textit{urn0} to \textit{urnM}), the probability of the special state with the whole of the \textit{N} marbles in the \textit{urn0} would be $(M+1)^{-N}$ \cite{Kao}. The corresponding so called average recurrence time, that is the time after which the system regains periodically (to arbitrary closeness) its initial state (Poincaré cycle) is then calculated from the Kac\textquotesingle s lemma \cite{Kac47,Falcioni}, taking the inverse of this probability: $m_{\left[N\to N\right]}= (M+1)^{N} m_1$, being now $m_1$ the time slice spent to force a marble to change urn.\footnote{See, e.g., \cite{Falcioni} for a proof of this. The value reported obviously depends on the peculiarity of the macroscopic state considered, namely the state of minimum probability (the \textit{N}-state or zero-entropy state).  Each macroscopic state has its own probability (known in the literature as stationary probability or Markov probability measure), ranging from a minimum in the \textit{N}-state to a maximum in the ${N\slash 2}$-state (the equilibrium state or maximum-entropy state). As the equilibrium state is reached, the probability to deviate from it is so much smaller than \textit{N} is bigger. In this near-to-equilibrium regime, the probability is well approximated by a Gaussian centered on $N\slash 2$.}

\section{Emergence of irreversible behavior} \label{sec9}

\begin{figure}[!ht] 
\centering
\includegraphics[width=1.0\columnwidth]{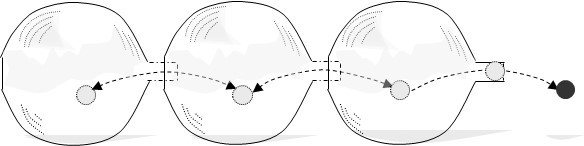}\caption{\small A chain of three bottles communicating through free flowing (two-way) openings between them. This is called a $C(3,1)$ assembly, denoting a chain of 3 bottles, with one forward opening and one backward opening ($M = 1$) each (except the first bottle that has only one opening). If the marble takes one hour to leave the first bottle, it will take about $3^2=9$ hours of shaking to get the marble out of the bottle chain.}\label{fig:image01}
\end{figure}

In the Marbles and Bottles-or Boxes model subject of this paper we have seen that, in the simplest case of a single two-way connector joining the bottles (figure \ref{fig:image01}), the number of bottles reached on average by the marble at a given time increases with the square root of the same time: $n\varpropto\sqrt{m}$, as expected for unidimensional random walks.\footnote{In the unidirectional forward motion without regression the growth would be linear: $n\varpropto m$.}

\begin{figure}[!ht]  
\centering
\includegraphics[width=1.0\columnwidth]{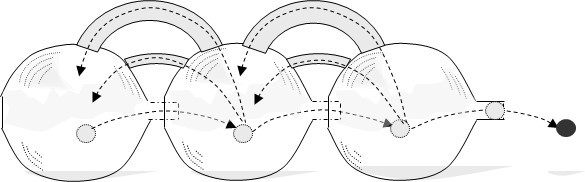}\caption{\small A $C(3,2)$ assembly: three bottles with two backward connectors ruled by one-way valves (always with the exception of Bottle 1). It will take an average of 19 hours of uninterrupted shaking for the marble to escape. (The enlarging of the hole of Bottle 1 required to make the first bottle equal to the rest is not shown).}\label{fig:image02}
\end{figure}

\begin{figure}[!ht] 
\centering
\includegraphics[width=1.0\columnwidth]{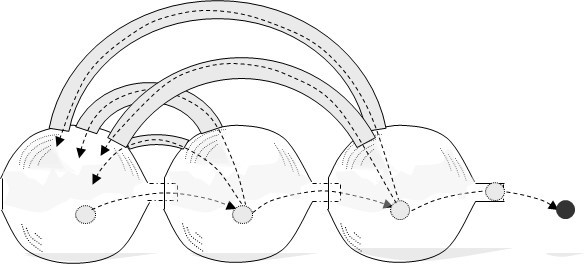}\caption{\small A variant of $C(3,2)$ with one-way tubes bringing directly the marble back into the first bottle. The marble escapes after 39 hours of enduring shake. (Additional tubes connecting Bottle 1 to itself, in order to make it equal to the rest, are not shown).}\label{fig:image03}
\end{figure}

The introduction of additional one-way backward connectors, as in figures \ref{fig:image02} and \ref{fig:image03}, leads to an increase in disorder in the path of the marble. Now, as the number of backward connectors increases, the marble is more likely to undergo regression; so, on average, it would take longer to reach a given number of chain nodes. This number grows very slowly with time, in fact more slowly than any growing function described by a power law. We found that for $M>1$ (and very long chains) the relationship between the length of the chain assembly and the expected elapsed (absorption\slash escape) time (or between a given node number and the corresponding average first-passage time) exhibits a logarithmic pattern. 


\begin{figure}[!ht] 
\centering
\includegraphics[width=1.0\columnwidth]{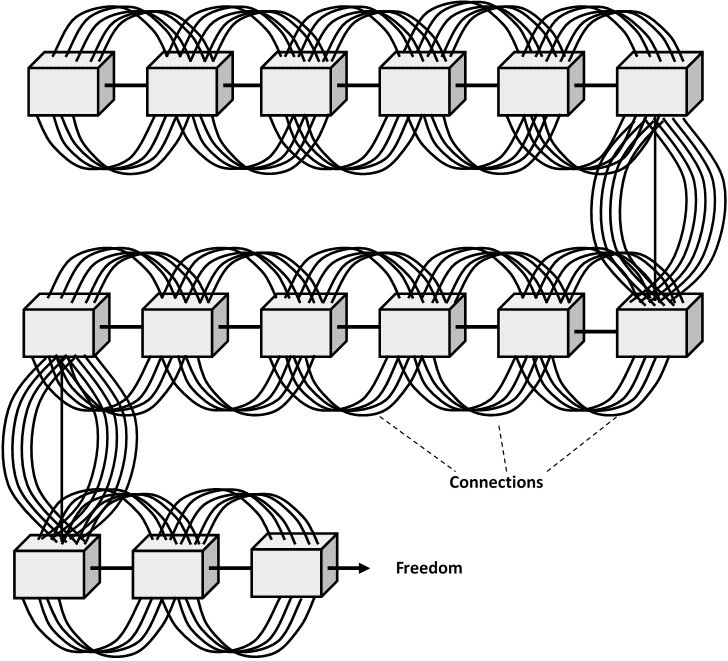}\caption{\small A $C(15,9)$ assembly, consisting of 15 boxes with 9 backward connectors each, represents about 6.6 billion years, far beyond the age of the solar system.}\label{fig:image04}
\end{figure}

Using either equation \eqref{eq:eps2.1} or its avatar \eqref{al5}, valid for suitably long chains (say $n > 10$), students can be encouraged to draw diagrams representing large time stretches, by varying the number of retrograde connections (represented by arcs) and \slash or the number of bottles-or boxes in the chain assembly. Some pictorial examples of such diagrams have been reported by Pickover, who also introduced -- in order to facilitate the discussion of the startling characteristics of these chains -- the symbol $C(n,M)$ to represent a generic chain assembly with \textit{n} chambers and \textit{M} backward connectors between them \cite{Pick02}.\footnote{Bear in mind that there is always one forward connector per node.} Each connector is represented by a line. Using Excel, you can easily map typical times of ``improbable'' processes, as the escape of a walker from such an intricate maze as that of figure \ref{fig:image04}.

It was apparently Smoluchowski (cited by Kac in \cite{Kac47}) who advanced the rule that a process started in a state with long recurrence time (that is - roughly speaking - the time to wait, on average, for the same state to recur) will appear as irreversible. On the other hand, a short mean recurrence time makes it meaningless to speak about irreversibility. Other different notions of (ir)reversibility and recurrence, and of their interrelationships, have characterized the many different formulations of classical thermodynamics in the last two centuries, leading to non-univocal meanings and sometimes confusing expositions of the same concepts\cite{Haddad}. Entering such subtle distinctions, however, is beyond the scope of this paper.

Marbles and Bottles-or Boxes models described in this paper are suitable for illustrating ``essentially'' irreversible processes because they are nothing more than absorbing chains of finite size, and all absorbing chains are actually not recurrent, being their exit probability (i.e. the probability that the walker eventually terminates at a particular node corresponding to an absorbing state) equal to $1$ \cite{KSK}. Ultimately, this non-recurring behavior has to be traced back to the eventual removal of the walker as it hits the absorbing barrier, so that the system can not be longer considered insulated. Differently, and perhaps surprisingly, infinite non-absorbing chains can exhibit recurrence, transience or ergodicity under sufficient conditions, while random walks on (finite) circular paths have the less restrictive recurrence conditions \cite{Denny,Ching}. The Sisyphus random walk considered in section \ref{sec7} is recurrent and ergodic on an infinite chain, as intuitively expected considering that the reset mechanism will prevent the walker from being driven too far off from the origin. Every point is reached infinitely often and the mean recurrence time is given by:

\begin{equation} \label{al12}
m_{\left[n\to n\right]}=\frac{(M+1)^n}{M}(M+1)m_1 
\end{equation}

(Cfr., for a proof, eq. (27) and eq. (11) in ref. \cite{Montero}). Note that equation \eqref{al12}, valid for an infinite chain, looks the same as the asymptotic expression of equation \eqref{al11} for the absorption time of the finite chain. It turns out, as intuitively expected, that the average time to surpass position \textit{n} and the mean recurrence time of same position both grow exponentially with the distance.

So, the emergence of irreversible behavior in systems modeled by finite - yet very large - absorbing chains is intended in a still stronger sense than the Smoluchowski\textquotesingle s criterion did, as the elapsed times involved are so long, compared to any ordinary experience, that the guess of the irreversibility (viz non recurrence) of the flow of marbles through these chains can be fully trusted. For example, a $C(15,9)$ assembly, that is an assemblage of fifteen chambers connected by one forward connector and nine backward connectors (see diagram of figure \ref{fig:image04}), represents a span of time of about $6.6\cdot 10^9$  years -- far beyond the age of the solar system (about $4.5 \cdot 10^9$ years) -- because the marble would spend that time to escape. With the addition of a single retrograde path, we realize a $C(15,10)$ assembly, which represents about $2.8 \cdot 10^{10}$  years or twice the age of the universe (about $1.4 \cdot 10 ^{10}$ years).
 
\section{Irreversibility Vs. Recurrence} \label{sec10}

To illustrate recurrence, the standard pedagogical approach contemplates urns containing numbered objects that are forced to change urn by some withdrawal mechanism. The relationship between the number \textit{N} of marbles returning on average to the zero-entropy state (the whole of the marbles in \textit{urn0}), the number \textit{M+1} of urns (\textit{2} in the basic version of the experiment), and the expected recurrence interval $m_{\left[N\to N\right]}$ takes the form $m_{\left[N\to N\right]}\slash m_1= (M+1)^N$, being $m_1$ the time slice spent to force a change of urn and make an observation. Note that the left-hand side of latter relation may be interpreted as the average number of observations (or time steps) expected until a recurrence incurs. This also means that for a given number of observations, the marbles ever-returning are limited on average to $N \le \log_{M+1}{\textrm{[number of observations]}}$. 

Unfortunately, a classroom activity on this subject is workable only for trivial values of \textit{N}. For example, assuming it would take 5 seconds to carry out one observation in the simplest two-urn experiment (extracting, moving, counting, etc.), the work required to observe the return of 10 marbles requires about 5,120 seconds, far beyond 1 hour. However, simple Excel simulations conducted by other authors have shown that all the main features of the Ehrenfest\textquotesingle s model can be tested in a few seconds \cite{Neri}.

Computer simulations suffer, however, serious limitations as the number of marbles begins to increase, due to the exponential growth of the computer\textquotesingle s time steps required: if we start, for example, with $N=100$ marbles in \textit{urn0}, then the expected return time is $2^{100} m_1$. Even if each simulation step required a billionth of a second, the entire run would take about $4 \cdot 10^{13}$ years to complete, or roughly 3,000 times the length of time that the Universe has existed thus far! Simulation runs with $N \cong 100$ (or higher) are therefore stopped at a very early stage, but even a limited number of run-steps will suffice to highlight the convergence towards equilibrium, which is attained rather quickly. For \textit{N} values within this order of magnitude, simulations give evidence that the amplitude of the fluctuations decreases with \textit{N}, so there is no hope of incurring a return. For a workable compromise, it would need to drop to $N \cong 40$ (or lower) to observe recurrence. 

Despite the above limitations, the simulations allow the following key points to be highlighted:
\begin{itemize}
\item the equilibrium state is the state of maximum probability;
\item the attractor character of the equilibrium state on systems far from it (A far-from-equilibrium system tends to evolve towards equilibrium);
\item the inverse process (spontaneous system transition from a more probable to a less probable state) is always possible, yet improbable.
\end{itemize} 
Concrete classroom experiences conducted in past years by other authors proved that two-urns simulations can support the appropriation of the concepts of statistical equilibrium, irreversibility, entropy, and unidirectionality of time\footnote{Simulations of this type were carried out without a computer directly with students: each student represents a particle in an initial state (\textit{urn0}), the teacher takes the name of a student at random and pushes him to move\ldots but  these experiments are (obviously) limited!} \cite{MarieJeanne}. 

Although with some differences, urn and random walk models have various similarities and share important characteristics. It was Kac, in 1947, who pointed out the equivalence between the Ehrenfest\textquotesingle s two-urn problem and the discrete random walk formulation of the Brownian motion of an elastically bound particle, when the excess over $N\slash 2$ of marbles in \textit{urn0} is interpreted as the displacement of the particle \cite{Kac47}.\footnote{Kac himself attributed to Schr\"odinger and Kohlrausch in 1926 the original insight about the connection between the two models (see \cite{Kac47}, p. 380).} 

\begin{figure}
\centering
\subfloat[Symmetric ($M = 1$) RW elapsed (blue) and RWR recurrence (red) times compared with two-urn recurrence time (dotted line, secondary scales).]{%
\resizebox*{1.0\columnwidth}{!}{\includegraphics[width=1.0\columnwidth]{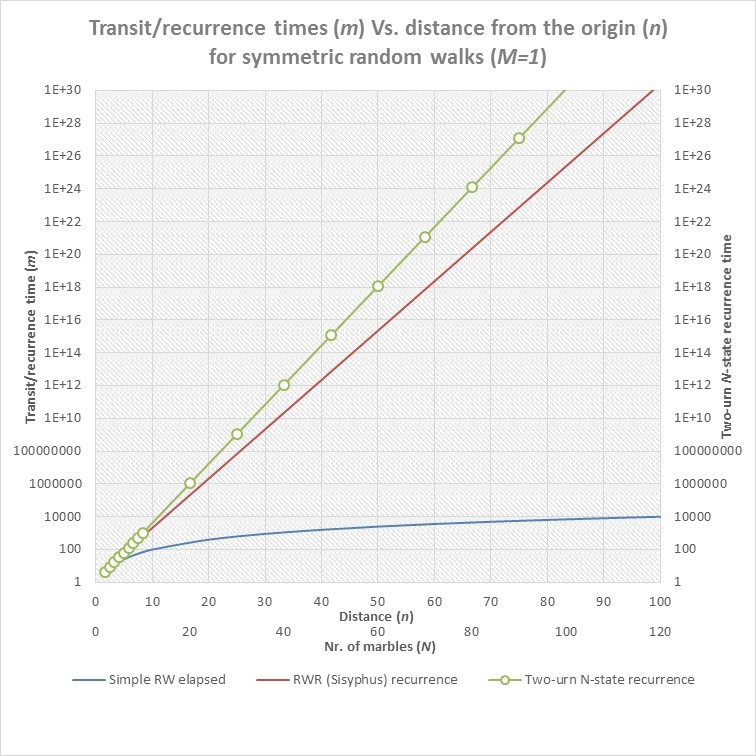}}}\hspace{5pt}
\subfloat[The same as (a) for the asymmetric case ($M > 1$). Data are plotted for \textit{M=2}. The marbles expand in two urns before returning to \textit{urn0}.]{%
\resizebox*{1.0\columnwidth}{!}{\includegraphics[width=1.0\columnwidth]{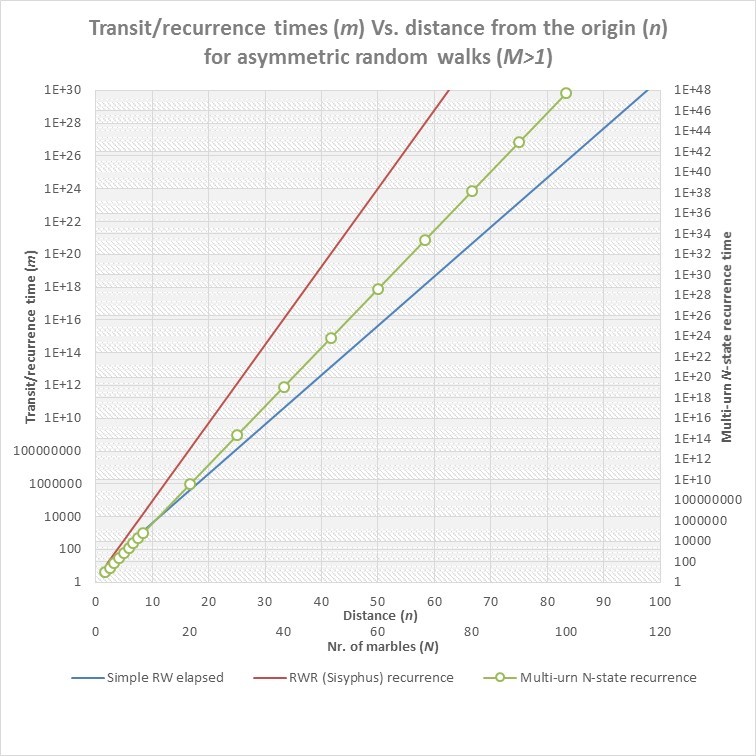}}}
\caption{Logarithmic plot of transit/recurrence time versus distance of the walker from origin for different systems.} \label{fig:graph01}
\end{figure}

\begin{figure}[!ht] 
\centering
\includegraphics[width=1.0\columnwidth]{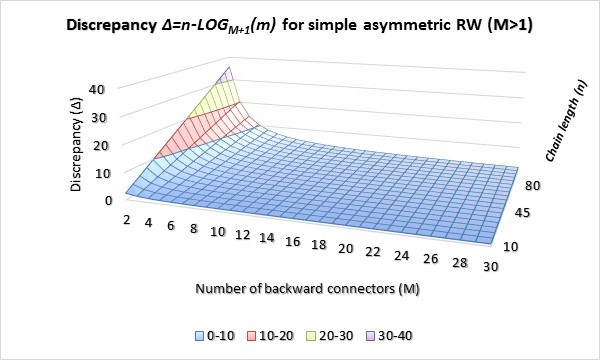}\caption{\small For the simple asymmetric random walk ($M > 1$), the discrepancy from a pure exponential growth ($\Delta=n-\log_{M+1}\textrm{[elapsed time]}$, vertical axis) is plotted versus the number of backward connectors (\textit{M}, horizontal axis) and of the chain length (\textit{n}, depth axis). The resulting surface rises steeply for high \textit{n} and flattens out for high \textit{M}.}\label{fig:graph04}
\end{figure}

With no doubt both models contain the same main message: the emerging discrepancy between irreversibility and recurrence observed at the macroscopic level, when the statistical behavior of the system takes relevance. To realize it, it is worth comparing characteristic ``diffusion'' times of the two models by running simple simulations. Using Excel we have carried out a sample mapping of elapsed times for random walkers transiting across unidimensional chains (typically non-recurrent) and recurrence times in urn-like and Sisyphus models. It turns out that all these characteristic times tend to be comparable with each other as the number \textit{N} of marbles initially in \textit{urn0} is interpreted as the distance \textit{n} covered by the walker, the random walk is increasingly asymmetrical (viz \textit{M} increasing) and the marbles expand in \textit{M} urns before returning to \textit{urn0}. Summary results of our simulations are plotted in figures \ref{fig:graph01} and \ref{fig:graph04} (where unit time steps are assumed).

\section{A physical argument: effusion of a low-pressure gas} \label{sec11}

Returning at last to the question at the bottom of section \ref{sec2}, concerning the passage from the experiment with marbles and bottles to the microscopic scale, let us consider, in place of marbles, particles of molecular size in motion at thermal velocities in sealed containers (imagine that there is a vacuum outside the container). If a pinhole large enough for particles to fit through is punctured in a wall of the container, a ``gas'' leakage from the pinhole is observed, with a characteristic effusion time.

Particle escape times from confined regions have been evaluated by various methods, either numerically – by means of classical mechanics (see \cite{DeLuca} for an example) – or by direct random walk Monte Carlo simulations of the effusive process \cite{Levin}. A rough idea of the typical time of a particle escape process can be attained through a parallel with low-pressure (say $<0.1$ Torr) effusion experiments with gases \cite{SGS}. The characteristic effusion time for air in a 1 liter container (in a vacuum) at room temperature (25\,\textcelsius) punctured by a 1 mm\textsuperscript{2} hole turns out to be about 8.6 s.\footnote{The characteristic time of effusion  (or ``relaxation time'') is given by $\tau=4V\slash A\bar{v}$, where: $V=$ Volume of the container, $A=$ Area of the pinhole, $\bar{v}=$ Average speed of molecules (467\,m\textfractionsolidus s for air at 25\,\textcelsius), provided that the pressure outside the pinhole is essentially zero (in practice $\lesssim 10^{-5}$ Torr); also $\bar{v}=\sqrt{8RT\slash \pi M}$, where $M=$ Molar mass (28.97\,g for air) \cite{SGS}.}

Even if one brings down the characteristic time $m_1$ to the time scale of seconds, thereby consistently alleviating the particle trapping effect in the relatively simple assemblies of figures from \ref{fig:image01} to \ref{fig:image03}, yet for chains of increasing length and complexity (like that of figure \ref{fig:image04}) the escape time quickly reaches values experimentally inaccessible.

\section{Conclusions} \label{sec12}
The models discussed in this paper can be a valuable educational tool, fostering a basic understanding of the statistical nature of the irreversible behavior of macroscopic systems. Urn-like models are traditionally considered advantageous for acquainting students with the concept of thermodynamic equilibrium and with the statistical origin of macroscopic irreversibility. On the other hand, the escape of a random walker out of a long chain of communicating compartments - similar to the escape from a maze - exemplifies in a more immediate sense an ``improbable'' process, possible in principle but that would require an unreachable amount of time in order to be actually experienced, close to - and in some sense stronger than - the Smoluchowski\textquotesingle s conception of irreversibility (irreversible process = non-recurrent initial state in any conceivable experiment).

Taken together, the two models give evidence that macroscopic irreversibility does not manifest itself at a fundamental level but as a result of statistical nature embodied in the macroscopic approach. Oppositely, the reversible character of phenomena, when examined at the microscopic level, appears to be ineluctable.\footnote{Be careful not to overemphasize reversibility at the micro level: examples were given of extremely simple, reversible, insulated systems that exhibit irreversible statistical behavior \cite{Hobson}.}

Another advantage of these models is that both are very intuitive and do not require that students have any prior knowledge of the subject. In addition, the number of variables to be understood is extremely limited: only one variable - the initial number \textit{N} of marbles in \textit{urn0} (or \textit{box1} in the Marble Game) -, or two - the number \textit{n} of chain nodes and the number \textit{M} of backward connectors - for the chain assembly.

Finally, both models give an occasion for rich disciplinary and interdisciplinary insights on the meaning of time, irreversibility, time arrow and other advanced topics. However, a dose of caution is needed. It should be kept in mind that if the level of subjects covered by these topics is very high, they cannot be developed in all types of  secondary schools and careful consideration must be paid to the skill level and interest of students.

\section*{Acknowledgments}
The authors are grateful for the valuable correspondence from Dr Clifford A. Pickover. Special thanks also go to Alessandra Hendry and Marina Maxia for careful reading of a prior version of the manuscript.

\end{document}